\documentstyle[epsfig]{basi}
\begin{document}
\title[NGC 1245]
{ NGC 1245 - an intermediate age open cluster}
\author[Annapurni Subramaniam]
{Annapurni Subramaniam $^1$ \thanks{e-mail: purni@iiap.ernet.in}\\
$^1$ Indian Institute of Astrophysics, II Block Koramangala, Bangalore 560 034, 
India}
\maketitle

\begin{abstract}
The B,V CCD photometry of 916 stars in the field of the high galactic
latitude intermediate age open cluster NGC 1245 is presented. The cluster parameters were
estimated with the help of V vs (B$-$V) colour magnitude diagram (CMD).
After correcting for intra-cluster reddening, the mean redding towards the
cluster was found to be 0.29$\pm$0.05 mag, with a distance modulus $(m-M)_0$ 
equal to 12.4$\pm$0.3 mag.
The cluster is found to be located at a distance of 3 Kpc. The isochrone fits using
Girardi et al. (2000) isochrones to the cluster CMD estimated an age of 890$\pm$100 Myr.
The synthetic CMDs obtained using Girardi et al. (2000) models incorporating
photometric errors and 10\% binary stars estimated an age of 1 Gyr for the cluster.
The luminosity function (LF) of the main-sequence 
shows dips, which might arise due to some known gaps in the main-sequence, including the
Bohm-Vitense gap.
The LF and integrated LF computed from the observed CMD is compared with those computed 
from the synthetic CMDs for five values of the mass function. The estimation of the
present day mass function indicates a flatter value compared to the Salpeter value,
though an accurate estimation is not possible due to the large dips found in the LF.
The apparent paucity of brighter stars seen near the cluster center is 
estimated to have a statistical significance of 2 $\sigma$. 
\\
\end{abstract}

\section{Introduction}
The open star cluster system of our Galaxy is one of the important constituent
of the disk of the Galaxy. The open star clusters are formed from molecular clouds
at the sites of star formation.  In general, the open star clusters are found 
very close to the plane of the galaxy, typical scale height being, 
100 -- 200 pc. Some open clusters are found to be located away from the disk
of the Galaxy, more than the scale height. 
NGC 1245 is an intermediate age star cluster situated close to the Perseus spiral
arm and is located 444 pc below the plane of the Galaxy. The location of this
cluster and its irregular appearance got us interested in this cluster.
NGC 1245 (RA = 3h 11.2m; Dec = +47 4) is located away from 
the center of the Galaxy (l=146.6; b=-8.9).
This intermediate age cluster might belong to the old disk population
with a larger scale height of about 500 pc. The intermediate age clusters
located far away from the galactic plane could also
throw some light about the old disk or the  thick disk population of the Galactic disk.

This cluster has been studied previously by Hoag et al. (1961) and 
Chincarini (1964). The cluster
finding chart presented by Hoag et al. (1961) shows that the cluster 
seems to lack stars
near the cluster center. Another study of this cluster based on CCD BV photometry
is done by Carraro \& Patat (1994). They obtain a colour excess E(B$-$V) = 0.26 mag
and an apparent distance modulus (m$-$M) = 13.20 mag. Wee \& Lee (1996) have carried
out Washington CCD photometry of this cluster and found that this cluster 
is not metal rich, given in the Lyng\aa\ (1987) catalogue.
A comparison with the photographic data given at the site WEBDA, 
indicates that there may be
calibration problems in Carraro \& Patat data. This has also been indicated by 
Wee \& Lee (1996). Therefore, we present a new set of CCD photometry 
for the stars in the region of NGC 1245.
The cluster colour-magnitude
diagrams (CMDs) are used to estimate the reddening, age and distance to the
cluster. The synthetic CMDs are used to estimate luminosity functions in order
to compare with the observed ones and also to estimate the present day mass function
of the cluster.
\section{Observation and data analysis}

\begin{figure*}
\centering
\includegraphics[width=13cm]{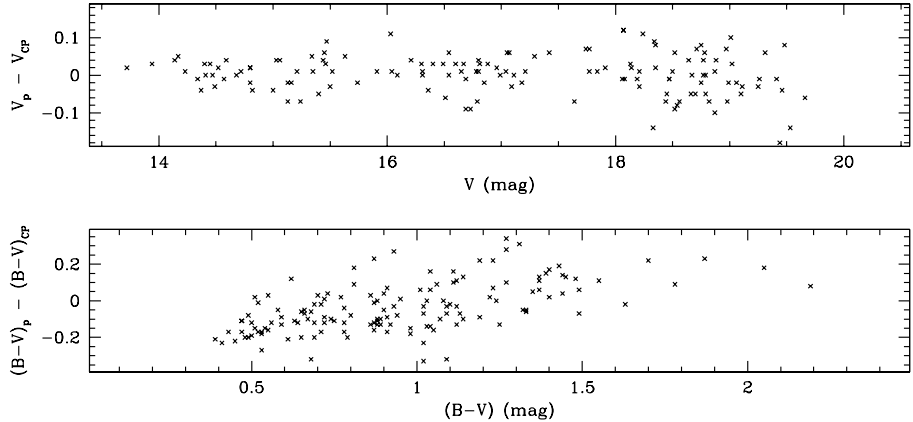}
\caption{The present data is compared with the data from Carraro \& Patat (1994)
and the results are presented here. Carraro \& Patat (1994) data is subtracted
from the present data and the difference is shown as a function of the present
colour/magnitude.}
\label{figure2}
\end{figure*}
The cluster was observed from the 2.34m Vainu Bappu Telescope at Kavalur, 
India during the 1-4 Jan 2000 observing run. The observational log is 
presented in Table 1. We used a 1K x 1K CCD camera at the prime focus,
with an image scale of 0."63 arcsec/pix.
The night of observation was not photometric and hence
we used the stars observed by Hoag et al. (1961) as calibration stars.
The CCD data were calibrated using IRAF data reduction package. The 
instrumental magnitudes were estimated using the DAOPHOT II package.
The zero point errors are of the order of 0.02 mag. The V magnitude and
(B$-$V) colour are estimated for 916 stars. 
The present data is compared with the CCD data
of Carraro \& Patat (1994) and the results are shown in Figure 1.
The differences are computed as present data minus Carraro \& Patat (1994) data,
and the difference is plotted against the respective
magnitude or colour. The average value of $\delta V$ is found to be 0.01 mag,
with $\sigma$ = 0.04 mag, and the average value of $\delta (B-V) $ is 0.04 mag,
with $\sigma$ = 0.02 mag. The colour plot shown in the bottom panel of figure 1 
shows that the colour difference has a colour dependence, indicating colour
calibration problems with  Carraro \& Patat (1994) data. 

\begin{table}
\caption{Journal of observation.}
\vspace{0.20cm}
\begin{tabular}{ccccc}
\hline
 Date    &    Filter&  UT &   Airmass& Exp. Time (Sec)\\
\hline
 Jan 02 2000 &   B &  14 21& 1.51&  60.0 \\ 
 Jan 02 2000 &   B &  14 23& 1.50& 300.0  \\
 Jan 02 2000 &   B &  14 32& 1.48&1200.0  \\
 Jan 02 2000 &   V &  13 45& 1.58&  60.0  \\
 Jan 02 2000 &   V &  13 48& 1.57& 300.0  \\
 Jan 02 2000 &   V &  13 58& 1.53& 900.0  \\
\hline
\end{tabular}
\end{table}

\section{Data incompleteness}
In general, the incompleteness in the photometric data is a function of stellar crowding and
stellar magnitudes. In the case of open clusters, the crowding is very minimal and
hence the incompleteness is mainly a function of the stellar magnitudes. 
We estimated the completeness of the photometric data in both B and V passbands
in the long exposure frames. The completeness in the data changes slightly from the
center of the frame to the periphery. Hence the data completeness for the cluster
and the field region are tabulated in table 2.
\begin{table}
\caption{ The data completeness in B and V pass bands in percentage for the cluster and field
regions are tabulated here.}
\vspace{0.20cm}
\begin{tabular}{cccccc}
\hline
B(mag)& cluster CF & field CF & V (mag) & cluster CF & filed CF \\
\hline
upto 18.0 & 100.0 & 100.0 & upto 17.0 & 100.0 & 100.0 \\
18.0 -- 19.0 & 99.0 & 100.0 & 17.0 -- 18.0 & 98.0 & 98.0 \\
19.0 -- 20.0 & 98.4 & 99.2 & 18.0 -- 19.0 & 96.0 & 97.0 \\
20.0 -- 21.0 & 94.0 & 94.0 & 19 -- 19.5 & 95.0 & 96.5 \\
\hline
\end{tabular}
\end{table}

\begin{figure*}
\centering
\includegraphics[width=13cm]{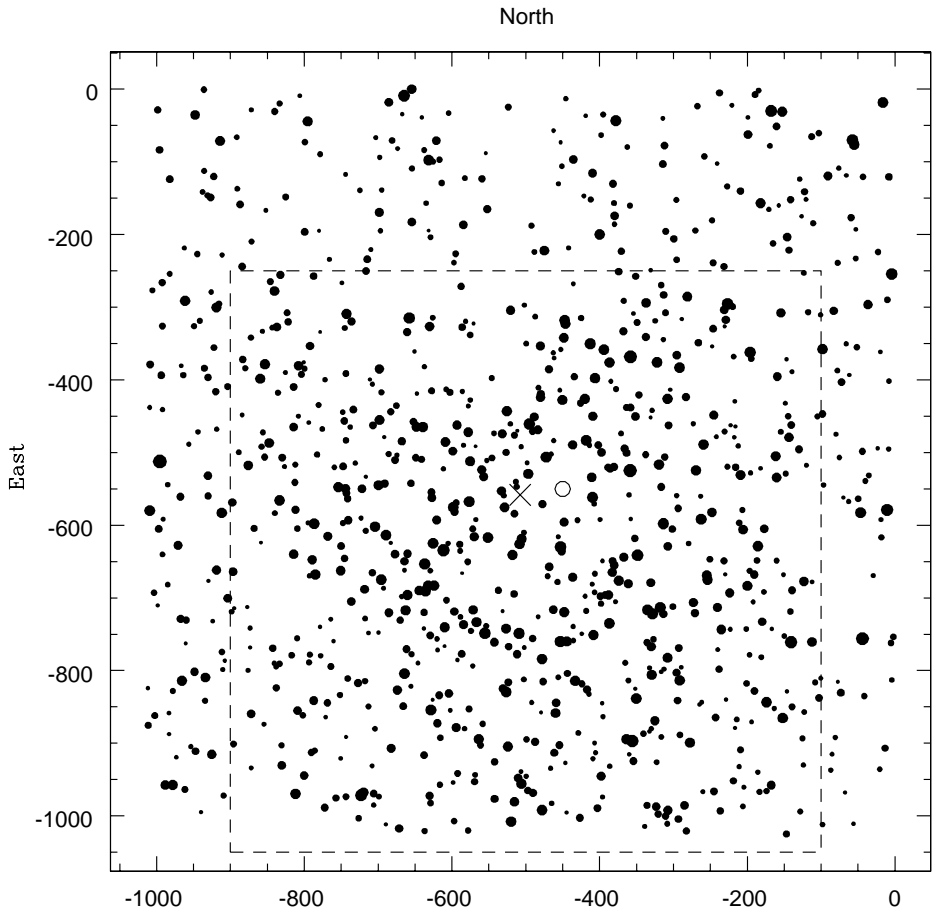}
\caption{The observed region of the cluster NGC 1245 is shown in the figure.
The cluster is assumed to be confined within the dashed lines.
The cluster center as found by Carraro
\& Patat (1994) is denoted by a big open circle. The center found in the
present analysis is denoted by cross.}
\label{figure3}
\end{figure*}

The observed region of the cluster is shown in Figure 2.
The cluster center as found by Carraro
\& Patat (1994) is denoted by a big open circle. We used this as the cluster
center and re-estimated the cluster center using all the observed stars. The
center thus found is marked by a cross. It can
be clearly seen from the cluster plot that the brighter cluster members are confined
within a radius of around 400 pixels. Thus 
 we have assumed that the
cluster occupies the area within the dashed lines and the region outside the
dashed box represents the field region. We estimate the field star
contamination in the cluster region using the data in this region. 

\section{Structure of the cluster}
From the cluster plot as  shown in Figure 2, it can be seen that
the cluster seems to lack stars near the center and also, the stars
seem to be distributed in an annulus. 
The cluster center as found by Carraro
\& Patat (1994) is denoted by a big open circle. We used this as the cluster
center and re-estimated the cluster center using all the observed stars. The
center thus found is marked by a cross. We estimate the radial density
profile (RDP) to study the radial density structure of the cluster.
The stellar density are estimated in radial bins from the center.
We computed the RDP for radial bins of 0.5 and 1 arcmin. RDP were estimated including all
stars and also for stars brighter than 17 mag. Two reasons for considering only
brighter stars: (a) to estimate the statistical significance of the apparent paucity of
bright stars near the cluster center and
(b) brighter star RDP could be used to identify the presence of mass segregation in the cluster.
When estimating the RDP for all stars, the incompleteness in the data is incorporated.
The radial density profiles
are shown in Figure 3, the density in the left panels is estimated using a radial bin of
1 arcmin and that in the right panels using 0.5 arcmin. The top panels show the RDP
computed using all the stars and the lower panels show the RDP computed using stars
brighter than 17.0 mag.

\begin{figure*}
\centering
\includegraphics[width=13cm]{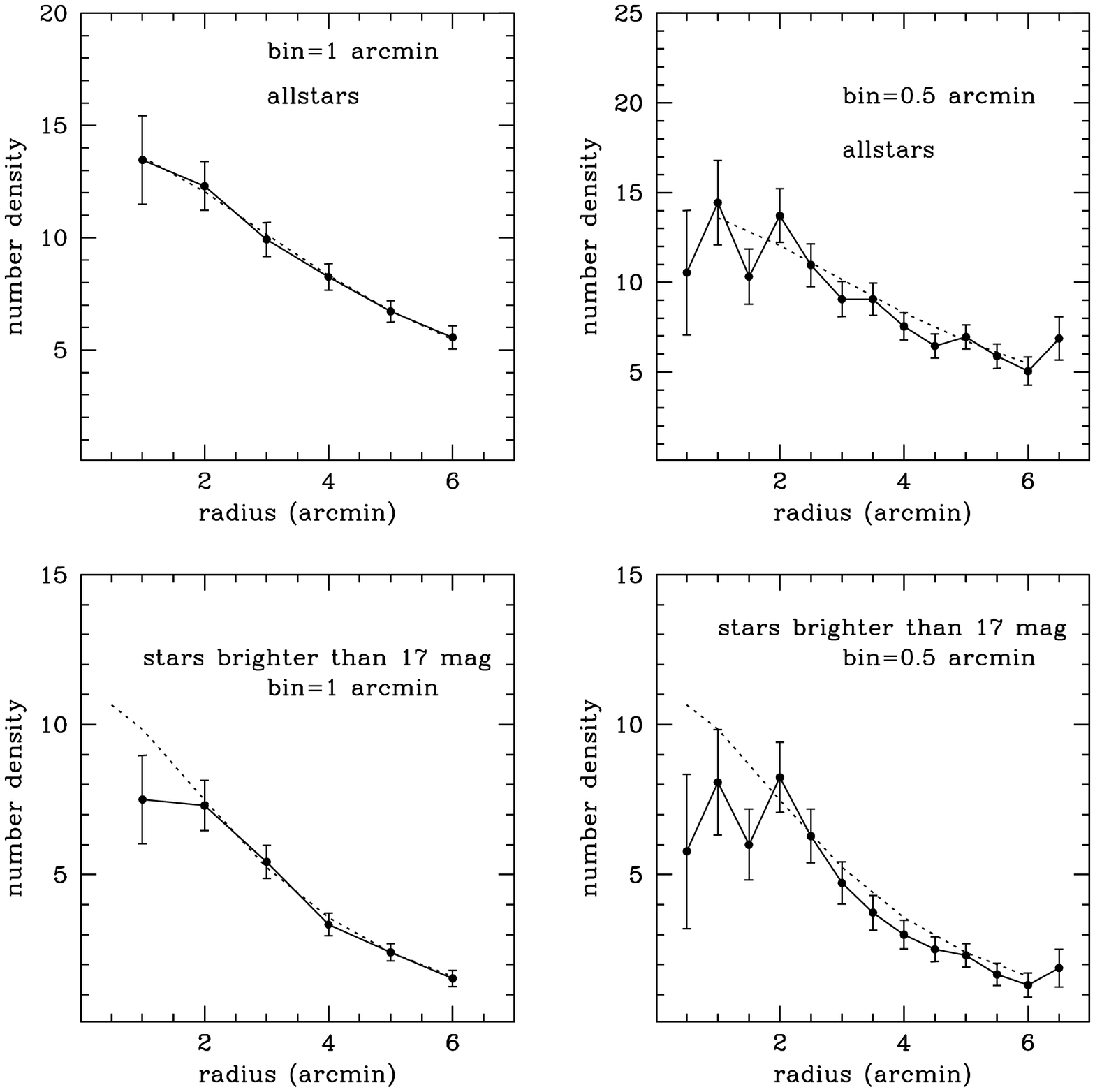}
\caption{The radial density of the cluster as a function of the
radius is shown in bold line. The left panel is for a radial bin of 1 arcmin
and the right panel for a radial bin of 0.5 arcmin. The dotted line shows the
fitted profile.}
\label{figure4}
\end{figure*}

It can be seen that the RDP in the top left panel is very smooth. This profile
is very similar to the RDP presented by Nilakshi et al. (2002), in their figure 1.
The top right panel shows a smooth profile outside 2 arcmin radius, but the
profile shows fluctuation near the center. 
This dip in the RDP is not seen by Nilakshi et al. (2002) as they consider
all stars with 1 arcmin radial bin and the dip is mainly seen in the RDP of
stars brighter than 17 mag. The RDP are fitted with the function
$ \rho(R) \propto f_0/(1+(R/R_0)^2)$, where $R_0$ is the core radius, at which the
density $\rho (R)$, becomes half of the central density, $f_0$.
These are also plotted in figure 3, shown as dotted lines.
In the case of top-left panel, the fit is extremely good with a $\chi^2$ value
of 0.15. The core radius is found to be 4.75 arcmin.
In the case of the lower-left panel, the fit to all the points turned out to be quite
unsatisfactory, with a $\chi^2$ value of 2.1. Instead, we fitted the function to the
points except the one at 1 arcmin, then the $\chi^2$ value was found to be 0.4.
This fit gives a core radius of 3.1 arcmin, which is much lower than the value obtained
when all stars were considered. If the fitted function were to extend to the inner radii,
at 1 arcmin, the observed density is found to be lower than the computed density.
The difference is about 1.6 $\sigma$. Hence RDP with 1arcmin radial bins for stars brighter
than 17 mag suggest that the central 2 arcmin has less number of brighter stars than 
what is expected. The RDP computed with half arcmin radial bins also show the fluctuation
in the number density of stars near the center. This is much more pronounced in the 
lower-left panel, where stars brighter than 17 mag are considered.
The RDP model profile, computed for the lower-left panel
is over plotted on the lower-right panel. The observed number density at 0.5 arcmin
radius is seen to be lower than the value expected by the fit, lower by 2 $\sigma$.
Hence this strengthens the point that the central density of bright stars is
likely to be less than what is expected.

The dip in the RDP could also be due to choice of center of
the cluster. The estimated cluster center is shown in figure 2. This center was
shifted by 50 pixels (0.5 arcmin) in all the four directions and the RDP
was re-estimated for stars brighter than 17 mag and with 0.5 arcmin bin.
The central dip in the profile was observed in three of the four 
RDP with shifted centers. 
Three RDPs with shifted centers showed a central density dips
which are more then 1 $\sigma$, indicating that the choice of the center is not 
the reason for the observed dip in the central density of stars brighter than 17 mag.

As this cluster is known to be an intermediate age cluster, the n-body relaxation
within the cluster would have resulted in the segregation of brighter stars
towards the center of the cluster. This would result in a steeper RDP
for brighter stars. This may be the reason for obtaining a lower value of
core radius for brighter stars. On the other hand, the bright star RDP is 
seen to be shallower than the fitted 
function in the central regions, though the statistical significance is 
not very high ( 2 $\sigma $). Also this result is independent of the errors in 
choosing the cluster center. Hence the apparent paucity of brighter stars near
near the cluster center has 2 $\sigma$ significance.

\section{Colour-Magnitude Diagram}
\begin{figure*}
\centering
\includegraphics[width=13cm]{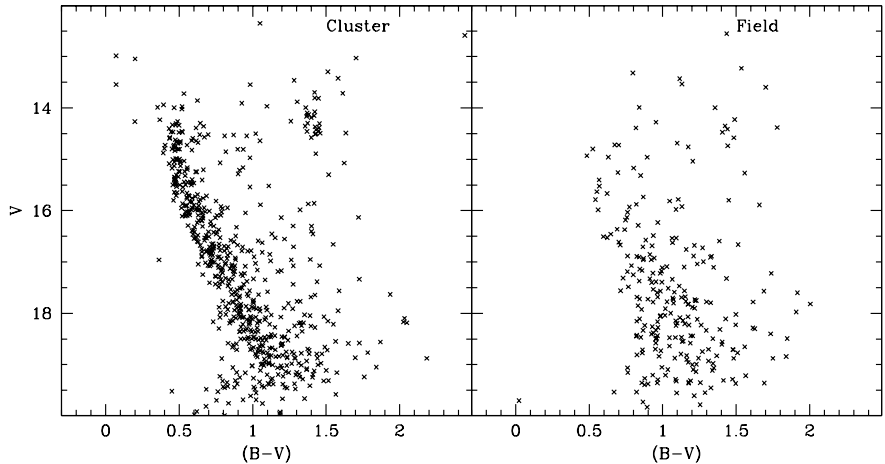}
\caption{The Colour Magnitude diagram of the cluster region and the
field region are shown here.}
\label{figure5}
\end{figure*}
\begin{figure*}
\centering
\includegraphics[width=13cm]{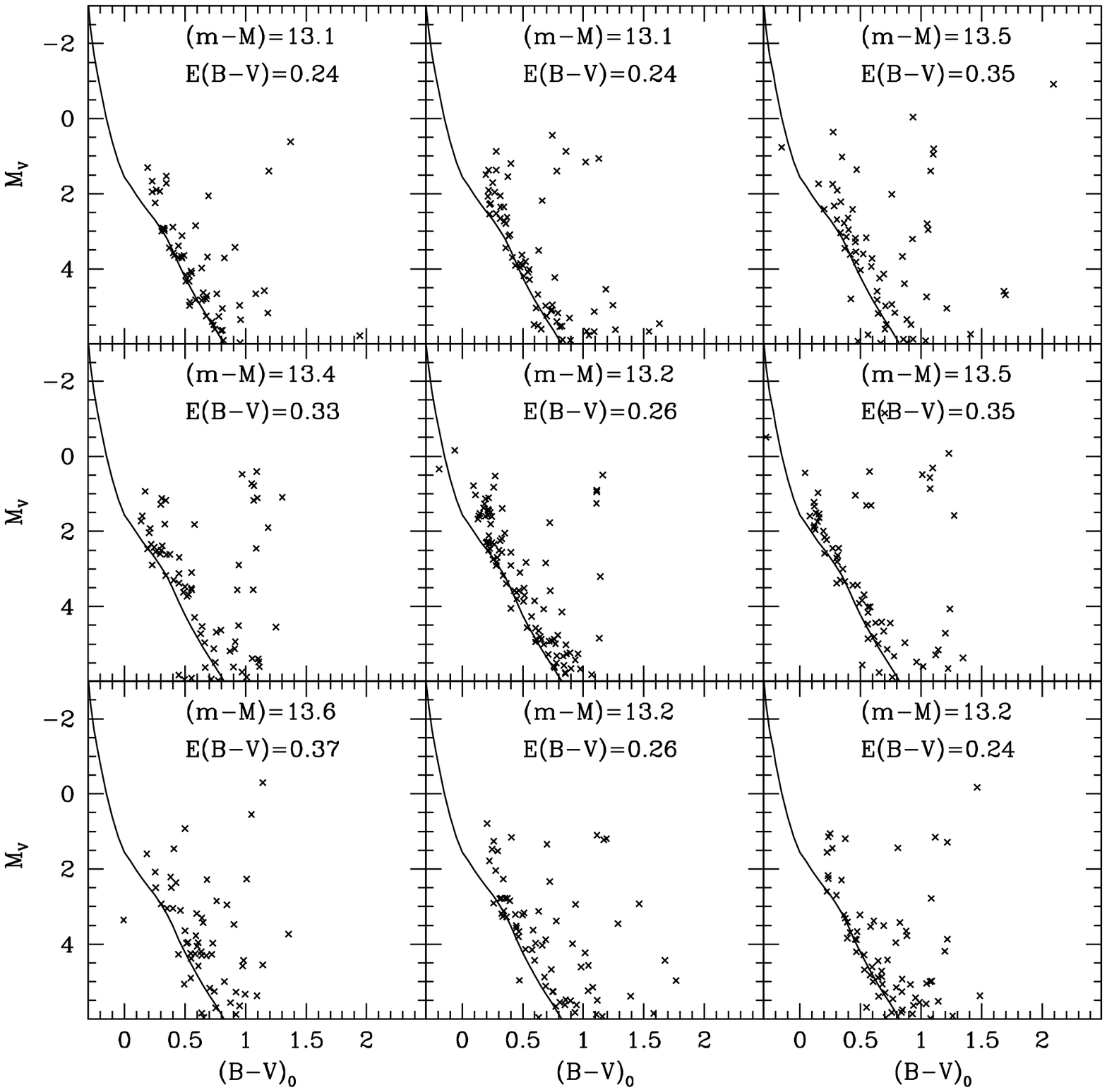}
\caption{The ZAMS fit to the 9 cluster regions are shown here. The
estimated reddening and distance modulus for each region are indicated.}
\label{figure6}
\end{figure*}
The parameters of the cluster  can be estimated using the CMD. 
The V vs (B$-$V) CMD is shown in Figure 4.
The left side shows the CMD for the cluster region with 659 stars and the right side
for the field region with 257 stars.  The cluster CMD spans between
V=13 -- 19.5 mag and a few red giants are seen at V= 14 mag and (B$-$V) = 1.4 mag.
A broad  main-sequence (MS) is seen, with a large number of stars
near the turn-off. One of the reasons for the scatter on the MS is the
presence of differential reddening across the face of the cluster, apart from the
presence of binaries. The differential reddening on the face of this cluster 
was noticed earlier by Carraro \& Patat (1994).
An estimate of this is required to obtain a more tight MS and thus to obtain
more accurate estimation of the cluster parameters.

\subsection{Differential reddening}
In order to estimate the amount of differential reddening, we divide the
cluster region into 9 equal regions. The CMDs for stars within
each region are constructed and fitted with ZAMS to estimate the individual
redding and distance modulus. The corresponding CMDs are
shown in figure 5, fitted with ZAMS. The estimated values of reddening
E(B$-$V) and the apparent distance moduli are also shown.

It can be seen that there is a significant amount of differential
reddening across the face of the cluster with the minimum 
and the maximum values of reddening being 0.24 and 0.37 mag respectively. 
It must be noted that the values estimated 
for each region may also be an average in itself. Further refinement in the 
estimate is not possible as the number of stars per region becomes too small 
to reliably estimate reddening.  Using the estimated individual
reddening per region, the stars in each region are de-reddened and are also
corrected for extinction. The cluster CMD after the above procedure is
seen to have reduced scatter in the MS. This CMD is used
for further analysis. The mean reddening towards the cluster is estimated
to be E(B$-$V)=0.29 $\pm$ 0.05 mag and the absolute distance modulus is 12.4$\pm0.3$ mag. 
This corresponds to a distance of
3020 pc, or 3 Kpc.
The error
involved in the estimation of distance modulus is 420 pc.
The estimates of reddening and distance obtained here are in excellent agreement
with the values obtained by Carraro \& Patat (1994).

\section{Cleaning the CMD}
The CMD after the removal of the intra-cluster reddening has the following
two defects. They are the presence of field stars and incompleteness in the
stellar data. In order to overcome these problems, we used a statistical method.
The method works as follows. 

First, the CMDs of the field and the cluster regions 
are divided into various small boxes. These boxes have a width 0.25 mag in V 
and 0.2 mag in (B$-$V). The number of stars in each box is estimated.
For each box, the data incompleteness values for the mean values of V and B 
magnitudes are estimated and the lower of the two values is adopted as the 
data incompleteness value for the box.
Then the number of stars required to be added in each box in 
order to make the data complete, is calculated. These stars are then distributed
randomly within the box. This procedure of adding extra stars in accordance with 
the incompleteness in the data in various regions of the CMD is done for both 
the cluster and the field CMDs. The area considered for field stars is less than
cluster area and the cluster area is found to be 1.4 times more than the field area. Hence
the number of stars in the field region needs to be multiplied by the above number to
be able to compare the field CMD directly with the cluster CMD. Hence the extra stars arising out
of the above factor is also randomly distributed within the boxes.

The next procedure is to remove the field stars from the
cluster CMD, using a technique called zapping technique. 
A very similar procedure is
used in the analysis of LMC clusters in Subramaniam \& Sagar (1995).
The cluster CMD contains cluster
stars as well as field stars and these field stars are required to be removed to 
obtain a CMD with only cluster stars.
In this method,
for each star in the field CMD, the nearest star on the cluster CMD is
removed. For a star in the field CMD, the same position on the cluster CMD is chosen and 
then the cluster CMD is searched for a star closest to this position by searching
within a V vs (B$-$V) box and increasing the size of the box until a star is found.
The range chosen to search for the star in the cluster CMD is a maximum of 
1 mag in V and 0.5 mag in (B$-$V). This procedure is repeated for all the stars
in the field CMD. Once all the thus identified field stars are removed from the cluster
CMD corresponding to the stars in the field CMD, the resulting CMD can  be considered 
to be devoid of
field stars. This CMD is presented in figure 6. 
\begin{figure*}
\centering
\includegraphics[width=8cm]{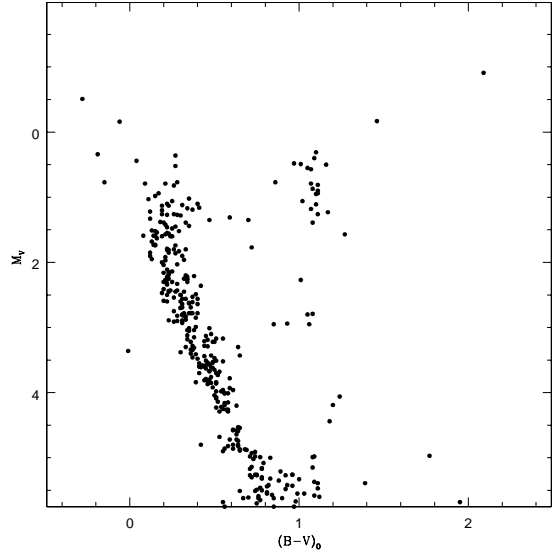}
\caption{The CMD corrected for differential reddening, data incompleteness
and field star contamination is shown here.}
\label{figure7}
\end{figure*}

As the procedure adopted here
is a statistical one, the stars which have the maximum probability to be a field star
in the cluster CMD are removed. Therefore the probability of the presence of field stars
in the final CMD is very less, though a few field stars still may be present.
A look at the CMD presented in figure 6 shows that the scatter seen at the top of
the CMD  remains the same, indicating that this truly is a cluster feature.
The MS is quite narrow and much refined when compared to the cluster CMD, shown in
figure 4. The noticeable feature on the MS is the clumpy distribution of stars along
the V magnitude and the sharp decrease in the number of stars near the limiting
magnitude of the present analysis, that is $M_V$ = 5.7 mag. The presence of a few
stars near $M_V \sim$ 3.0, 4.5 mag and $(B-V)_0 \sim$ 1.0 mag can be noticed and these stars
are most likely to be the left over field stars. 
On the whole this CMD can be assumed to consist only of cluster stars and hence this 
CMD is used for further analysis.

\section{Estimation of Age}
The CMD  which is corrected for intra-cluster reddening, data incompleteness and
field star contamination is used to estimate the age of the cluster.
We used the isochrones from Girardi et al. (2000) for age estimation.
Girardi et al. (2000) presented isochrones for Z= solar and Z=0.30 and also for
sub-solar values of Z. The metallicity  of the red giants in this cluster
as estimated by Wee \& Lee (1996) was [Fe/H] =  -0.04$\pm$0.05 dex. This
is very close to the solar value and hence we fit the solar metallicity isochrones.
Girardi et al. (2000) isochrones for sub-solar value of Z is for 0.008 which is
equivalent to [Fe/H] of -0.4 dex and this is very much less than that estimated 
for the cluster. Hence we use only the solar metallicity isochrones.
\begin{figure*}
\centering
\includegraphics[width=7cm]{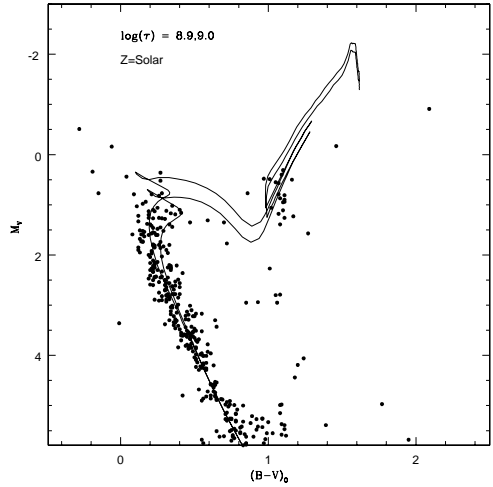}
\caption{The age of the cluster is estimated by fitting the isochrones from
Girardi et al. (2000). }
\label{figure8}
\end{figure*}
The isochrone fits for ages $log(\tau)$ = 8.9, 9.0 are shown 
 in figure 7.
Two stars are seen brighter and
redder than most of the  red giants. From the isochrone fit, it is difficult to consider them
as cluster members.  A few stars are seen brighter
than the MS turn-off, some of these might be candidate blue-stragglers. Hence
we estimate an age of 890 $\pm$ 100 Myr for the cluster.
The age estimated by Carraro \& Patat (1994) is 800 Myr, which
is within the error of the present estimation.

\section{Synthetic CMDs}
It is seen that the cluster CMD is better understood when compared with a simulated or synthetic
CMD made from the evolutionary models than with isochrones. A synthetic CMD
in general distributes stars along the isochrone corresponding to the age of the cluster,
in accordance with the
time-scale of each evolutionary state in the CMD. This feature is very useful  when comparing
the locations and number of stars in the red giant clump, sub-giants and stars at the tip
of the turn-off. This approach can also be used to estimate the
mass function of stars in the cluster.

We constructed synthetic CMDs using the theoretical 
stellar evolutionary models presented by Girardi et al. (2000). We used
solar metallicity models to create CMDs. The algorithm used to make 
synthetic CMDs are
presented in Subramaniam \& Sagar (1995) and also used in Subramaniam \& Sagar (1999).
We included the effects due to binaries with mass ratio between 0.75 -- 1.25 and also
the photometric errors in V and (B$-$V). Salpeter value for the mass function,
which is 2.35 (Salpeter 1955), is assumed in general. But this value could 
be changed to obtain the best fit when comparing the LFs.
A control parameter is required to create synthetic CMD, so that it can be
directly compared with the observed CMD. For this, one can either use the number of red giants, 
or the number of stars near the tip of the MS, below the turn-off. Since the red giant 
clump in this cluster is not very well defined and it
has less number of stars, we used the stars near the top of the MS for this purpose.
In the observed CMD, there are 17 stars between 15 -- 15.25 in V magnitude.
This is used as the control number. Initially we used a value of 30\% for the fraction
of binary stars
in the cluster, the resulting CMDs were seen to have too many stars near the evolved
part of the CMD, close to the binary isochrone path. Hence the binary fraction was reduced 
and a value
of 10\% was found to be satisfactory. Carraro\& Patat (1994) have assumed a value of 15\%
in their synthetic CMDs. 
\begin{figure*}
\centering
\includegraphics[width=8cm]{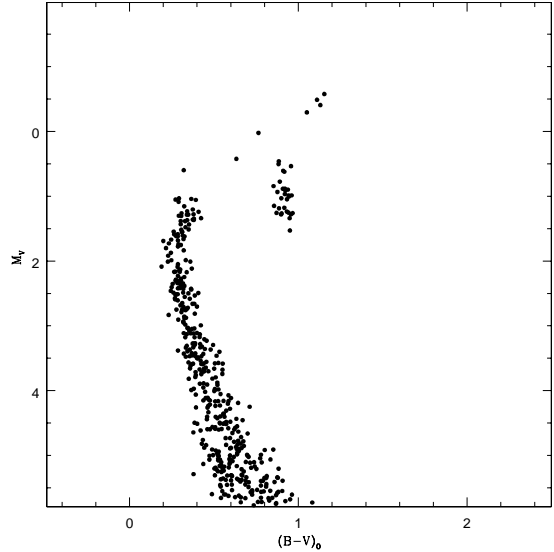}
\caption{The synthetic CMD is shown here. }
\label{figure9}
\end{figure*}

The synthetic CMD is shown
in figure 8. 
The value of mass function assumed is 2.35 (Salpeter value). 
The synthetic CMD was initially created for an
age of 890 Myr and it was found that the red giants were brighter than the ones on the
observed CMD. Therefore, the age was increased to a value of 1 Gyr and the red giant
magnitudes were found to match very well. Hence the synthetic CMD shown is for 
an age of 1 Gyr. It can be seen that the synthetic CMD mimics most of the 
features seen on the observed CMD. The vertical spread of the red giants is reproduced
beautifully. A few stars are created on the AGB branch, but we do not see any such stars
in the cluster CMD, except for the two bright and very red stars. A few stars are seen 
on the binary track, their number is again slightly more than what is observed, hence the
value of 10\% for the binary fraction could be considered as an upper limit.
The stars brighter than the MS turn-off are obviously not reproduced in the synthetic 
CMDs, so are the stars assumed to be field stars, located on to the right and fainter
end of the MS.

\begin{figure*}
\centering
\includegraphics[width=13cm]{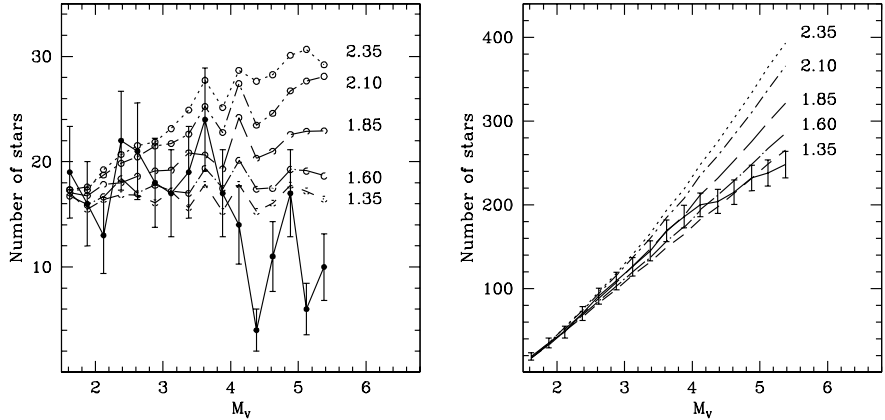}
\caption{The synthetic LFs are shown along with the observed ones 
(bold lines). The left panels show LF and the right panels show ILF.}
\label{figure10}
\end{figure*}

\section{Luminosity Functions}
As seen from the CMD of the cluster, the stellar distribution on the MS is very clumpy.
In order to quantify the above, we computed the MS luminosity function (LF) by counting
stars in magnitude bins on the $M_V$ axis. The observed cluster LF profiles are shown
in the left panels of figure 9 in bold line, with statistical error
at each point.
The observed LF profile  shows a lot of dips.
At $M_V$ = 2.1, 3.1 and 4.4 magnitudes, the profile dips considerably.
Around $M_V$ = 4.4, the number of cluster stars are found to 
be very minimal. 
The LF for a normal cluster shows an increase in the number of stars
per magnitude bin from brighter magnitudes to fainter magnitudes, and this is not seen 
in the LF profile of NGC 1245. This is on and above the dips seen in the profile.
To assess this point, we also computed the integrated
luminosity function (ILF), which is cumulative in nature and hence hides smaller fluctuations. 
The observed ILF profile is shown in the right panel of figure 9 as bold line
marked with statistical error at each point. A closer
look at the ILF profile reveals that slight kinks exist at the same magnitude levels as seen
in the LF profile. The ILF profile flattens for magnitudes fainter than $M_V \ge 4.0$ mag,
which corresponds to the large drop in the number of stars as seen in the
LF profile.

In order to compare the LF with those predicted by stellar evolutionary model,
 we have computed the
LFs from synthetic CMDs. The LF and ILF computed from the synthetic CMDs
are plotted in figure 9. These are shown as dotted, dashed and long dashed lines,
corresponding to five different values of the mass function. It can be seen that the number 
of stars
involved is very small, including the number of control stars used to create synthetic CMDs.
This results in high statistical fluctuations in the number of stars created, which
results in equally high fluctuation in the computed LFs.
Therefore, to compute reliable LFs, we have taken an average of 50 runs of the synthetic CMD
and this is assumed to
reduce the statistical fluctuation. As indicated by the slope of the
LF, CMDs were created for Salpeter value of the mass function and for values shallower 
than this.  Hence the CMDs were created for
the mass function values of 2.35, 2.10, 1.85, 1.60 and 1.35.  The value of the
mass function is indicated against each LF profile computed from synthetic CMDs.

The dips seen in the observed LF profile could be due to the gaps in the MS.
We see that at $M_V$ = 2.1, no gap is noticed previously in the cluster MS.
A slight dip in the simulated LFs is seen at this point, as also seen in the
synthetic CMD shown in figure 8.
On the other hand, at $M_V$ = 3.1, and $(B-V)_0$ = 0.33, there is a well-known
gap called the Bohm-Vitense gap. In some clusters, this is found at a value of 
$(B-V)_0$ of 0.28 (Kjeldsen \& Frandsen 1991), but recently some clusters were 
reported to have seen this gap
at $(B-V)_0$ of 0.33, like Hyades (de Bruijne et al. 2000 ). 
It is interesting to note that the simulated
LFs also show a dip at this magnitude. The most significant gap
at $M_V$ = 4.4 mag and $(B-V)_0$ = 0.55 is also very well seen in the observed CMD (figure 6).
In fact this gap is also seen in some old clusters like, NGC 6143, IC 4651, NGC 752 
(Kjeldsen \& Frandsen 1991). This luminosity corresponds to a mass of 1.15 $M_\odot$. This gap is
seen in the Hyades MS also, at $M_V$ $\sim$ 4.0 and $(B-V)$ = 0.48 and this
is mentioned as the second Bohm-Vitense gap (de Bruijne et al. 2000).
It is of interest to note that this gap is seen in the simulated LF.
Hence the point to be noted is that there are three significant dips seen in the
cluster LF, which are more or less reproduced in the simulated LFs. Therefore, there
is a good chance that the gaps and hence the dips in the cluster LF are real features.
 It is seen
that the second Bohm-Vitense gap is more prominent than the first one in this cluster.

The mass function of the cluster can be computed from the observed LF. In this cluster,
due to the dips present in the LF, an estimate of the mass function was not possible.
The standard plot of log(M) vs log(N) shows very large scatter and hence a straight line
fit to the data points was not possible.
Hence we used the ILF to estimate the mass function by comparing it with the synthetic ILF.
The magnitude range considered here is between $M_V$ = 1.5 -- 5.8 magnitude. This
range corresponds to a mass range of 2.2 -- 0.9 $M_\odot$. The lower mass range is
limited by incompleteness in the data. 
From the figure it can be seen that the profile corresponding to
the mass function value of 1.85 seems to fit the ILF profile better. The corresponding
LF profile can also be seen to take the mean values of the observed  LF. It should also
be noted that, the ILF profiles for the other mass function values do not deviate
very much from the observed profile at brighter magnitudes and the difference is seen 
mostly at the fainter ends.  The observed ILF profile flattens out from the x=1.85 profile
and almost touches x=1.35 profile at the fainter end at $M_V \ge 4.0$ mag. 
If we ignore this large dip at the end of the ILF, and consider the range 
$M_V$ = 1.5 -- 4.0 mag, then
the mass function value of 1.85$\pm$0.3 could be considered to fit the profile. 
The error in the mass function is estimated from the statistical errors 
in the observed profile, which happens to be within the nearby profiles corresponding to
the values, 2.1 and 1.6, of the mass function slope. The effect of inclusion of the fainter
stars would be to adopt a more flatter value of the mass function.
Though the present estimate shows that 
the cluster possibly has a flatter mass function slope, an accurate estimate is not possible
due to the large uncertainty seen in the fit.

Carraro \& Patat (1994)
found that the Salpeter value of mass function fitted the cluster ILF.
The present analysis uses much more
number of stars in the cluster than Carraro \& Patat (1994). 
Since the cluster is found to have stars distributed spatially 
in an irregular clumps, 
a subsection of this cluster might not represent a true cluster sample.
Hence, the present sample could be considered to be more complete. 
We shall discuss the possible reasons for a flatter present day mass function
in the section for discussion. 

\section{Discussion}

The estimation of cluster reddening and distance agrees quite well with those
estimated by Carraro \& Patat (1994).  The cluster is found to be 1 Gyr old, by the
synthetic CMD analysis, while Carraro \& Patat (1994) found the cluster to be 
slightly younger than this. 
Wee \& Lee (1996) estimated an age of 1.1$\pm$0.1 Gyr.
The estimation of age by the synthetic CMD method is
more reliable as it also considers the time-scale of evolution at each point on the 
CMD. Hence 1 Gyr could be considered as a more reliable age estimate.

The synthetic red giants are
seen to be bluer than the cluster red giants. This is unlikely to arise from the error
in choosing the cluster metallicity, as these red giants were found to be
very close to solar metallicity. 
This problem may be due to the mixing length considered in the 
convective envelope of red giants in the model calculations, or due to error in
colour-temperature relations.

The RDP is found to show that the cluster has a small deficiency of bright
stars near the center of the cluster, which basically endorses the visual perception.
As this cluster is 1 Gyr old, one can expect mass segregation to be present in the cluster. 
The core radius is estimated to be 4.75 arcmin when all stars were considered
and 3.1 arcmin for stars brighter than 17.0 mag. For a distance of 3 Kpc,
these correspond to 4.2 pc and 2.7 pc respectively. These values are higher than that
estimated by Nilakshi et al. (2002).
A smaller value for the cluster core for brighter stars,
as estimated from RDP would probably point towards the presence of mass segregation.

In general, any star cluster in the Galaxy, experiences the tidal force of the Galaxy, this
together with the n-body relaxation of the cluster, results in the cluster losing the low
mass stars to the Galactic field. This process is 
accelerated  due to many factors, such as
dynamical friction, encounters with molecular clouds, star clusters etc.
Hence over a time scale the number of low mass stars
in the cluster decreases. Being an intermediate age cluster, this cluster 
is likely to have lost stars at the
low mass end. The lost stars could also be present in the corona of the cluster.
Recently Nilakshi et al. (2002) found that clusters in general have large corona and
and the corona has a significant fraction of the cluster stars. The cluster area considered
here is not very big and a region adjacent to the cluster region is considered as the field region.
As mostly the low mass stars are spread out in the corona, the above two selection
criteria  could arise in under sampling and over subtracting stars at the low mass end.
The flattening of the ILF at fainter magnitudes could be due to this.
The present day mass function of the cluster is estimated to be 1.85, for the brighter end
and it is flatter than the Salpeter Value.
One needs to see whether this value changes  significantly
if stars in the corona are also included. 

\section{Results}
The summary and results of the present analysis are as follows:
\begin{description}
\item[1.] B,V CCD photometry of 916 stars are presented near the open cluster NGC 1245.
\item[2.] The cluster is estimated to be located at a distance of 3 Kpc, with an
average reddening of E(B$-$V), 0.29$\pm$0.05 mag.
\item[3.] The age of the cluster is estimated to be 1$\pm0.1$ Gyr using the models from
Girardi et al. (2000) and synthetic CMDs.
\item[4.]The LF of the main-sequence 
shows dips, which might arise due to some known gaps in the main-sequence, including the
Bohm-Vitense gap.
\item[5.] The cluster is found to have a flatter present day mass function 
compared to the Salpeter value of 2.35.
\item[6.] The apparent paucity of brighter stars seen near the cluster center is 
estimated to have a statistical significance of 2 $\sigma$. 
\end{description}

\section*{Acknowledgments}
I thank the referee for comments and suggestions which improved the paper.


\begin{thebibliography}{}
\bibitem{cp94}
Carraro, G., Patat, F., 1994, A\&A, 289, 397
\bibitem{c64}
Chincarini, G., 1964, Mem.S.A.It., 35, 2
\bibitem{}
de Bruijne, J.H.J., Hoogerwerf, R., de Zeeuw, P.T., 2000, ApJL, 544, L65
\bibitem{}
Girardi L., Bressan A., Bertelli G., Chiosi C., A\&AS, 141, 371
\bibitem{}
Kjeldsen, H., Frandsen,S., 1991, A\&AS, 87, 199
\bibitem{h61}
Hoag, A.A., Johnson, H.L., Iriarte, B., Mitchell, R.I., Hallam, K.L.,
Shapless, S., 1961, Publ. US Naval Obs., 17, 347
\bibitem{L87}
Lyng\aa, G., 1987, Catalog of Open Star Cluster Data (Strasbourg: CDS)
\bibitem{}
Nilakshi, Sagar, R., Pandey, A.K., Mohan, V., 2002, A\&A, 383, 153
\bibitem{s55}
Salpeter, E.E., 1955, ApJ, 121, 161
\bibitem{ss99}
Subramaniam, A., Sagar, R., 1999, AJ, 117, 937
\bibitem{ss95}
Subramaniam, A., Sagar, R., 1995, A\&A, 297, 695
\bibitem{}
Wee, S.O., Lee, M.G., 1996, JKAS, 29,181
\end{thebibliography}
\end{document}